\documentclass{article}

\usepackage{arxiv}

\usepackage[utf8]{inputenc} 
\usepackage[T1]{fontenc}    
\usepackage{hyperref}       
\usepackage{url}            
\usepackage{booktabs}       
\usepackage{amsfonts}       
\usepackage{nicefrac}       
\usepackage{microtype}      
\usepackage{lipsum}
\usepackage{graphicx}
\usepackage{color}
\usepackage{amsmath} 
\usepackage{multirow} 
\graphicspath{ {./images/} }

\title{Optimizing power by selective IP card shutdown using transport slicing}

\author{
 Alfonso Sánchez-Macián \\
  Department of Telematic Engineering\\ 
  Universidad Carlos III de Madrid\\
  Leganés, Spain \\
     \And
 \'Oscar {González de Dios} \\
  Telefónica Innovación Digital\\
  Madrid, Spain \\
   \And
 José Alberto Hernández \\
  Department of Telematic Engineering\\ 
  Universidad Carlos III de Madrid\\
  Leganés, Spain \\
  \And
   Liesbeth Roelens \\
  Telefónica Innovación Digital\\
  Madrid, Spain \\
   \And
    Pablo {Armingol Robles} \\
  Telefónica Innovación Digital\\
  Madrid, Spain \\
   \And
    Juan Pedro Fernadez-Palacios \\
  Telefónica Innovación Digital\\
  Madrid, Spain \\
   \And
  Ramón Casellas \\
  CCTC/CERCA\\
  Castelldefels, Spain \\
     \And
  Filippo Cugini \\
  CNIT\\
  Pisa, Italy 
}

\begin{document}
\maketitle
\begin{abstract}
The increasing energy demands of upcoming sixth-generation (6G) mobile networks and networks supporting AI applications pose significant challenges for  network operators in terms of operational costs and environmental impact. To address these challenges, this paper proposes a novel IP-based network slicing strategy that optimizes energy efficiency through a dual-slice approach. The proposed solution consists of a Day Slice, designed to meet high-performance requirements during peak traffic hours, and a Night Slice, optimized for energy savings by deactivating excess line-cards in card-based routers during periods of low traffic demand. The traffic is switched between the Day and Night Slices at predefined times, assuming appropriate traffic engineering mechanisms are in place to minimize disruption and support session continuity. We apply Pareto-based evolutionary algorithms (NSGA-II, CTAEA, and AGE-MOEA) to jointly optimize energy consumption and latency.
Experiments conducted on the SNDlib \texttt{india35} topology demonstrate that multi-objective optimization can deactivate over 40\% of line cards during low-traffic periods, providing significant energy savings while maintaining acceptable performance. Additionally, a multi-service extension using AGE-MOEA introduces differentiated QoS constraints, maintaining latency below 7 ms for premium traffic while preserving substantial energy savings.
\end{abstract}


\section{Introduction}
\textcolor{black}{Network slicing, as developed primarily by 3GPP, enables different groups of customers (e.g., vertical industries) to experience logically isolated, dedicated network services over a shared physical infrastructure, where each service is tailored to meet specific performance, security, and resource requirements.}
Similarly, in the IETF framework (\cite{farrel2024framework}), an IETF Network Slice is defined as an end-to-end logical network with a set of slice service objectives (such as bandwidth, delay, jitter, etc.) delivered through the underlying IP/MPLS technologies. Each slice operates as an independent logical network with its own topology, resources and unique configurations for bandwidth, latency, security, and routing policies, all of them on a shared physical infrastructure. IP Network Slicing allows the segmentation of an IP network into multiple virtual slices, each tailored for specific use cases or service requirements. This approach is particularly useful in 5G and beyond networks, cloud environments, and enterprise networking, where different applications or users require different levels of performance, security, or isolation.
In essence, slicing extends traditional network virtualization by also guaranteeing differentiated Service Level Agreements (SLAs) such as latency, bandwidth, reliability, or other Quality of Service (QoS) metrics for each slice (\cite{s23136053}).

This allows network operators to meet diverse requirements on the same IP network, e.g. one slice can be optimized for ultra-low latency (suitable for real-time applications or autonomous vehicles), while another slice is configured for high bandwidth (for video streaming or large data transfers). The fundamental idea is to transform a single physical IP network into multiple coexisting logical networks, each with customized routing and resource allocation to fulfill specific performance criteria, without having to build separate physical networks.

Backbone networks are typically dimensioned based on peak or busy-hour traffic demand. This guarantees performance under worst-case conditions. However, traffic follows stable daily patterns, with significantly lower demand during off-peak periods. This gap creates an opportunity for temporary energy savings.

Network slicing benefits from the development of a transport \textcolor{black}{Software-defined Networking (SDN)} architecture enabling network programmability through open and vendor-agnostic APIs (Application Programming Interfaces). A telco operator could offer APIs and network slices with specific Service Level Agreements (SLA) to other service providers.  

Network planning and operation processes require considering energy aspects. The main goal is to create networks that are greener and more sustainable, with energy consumption and carbon awareness serving as parameters. The objective of energy-aware transport slicing is to reduce power consumption without compromising the operational integrity, reliability, or resilience of the network. 

This paper proposes the creation of different network slices in which energy efficiency is prioritized over network \textcolor{black}{performance} characteristics, with the option of switching between them depending on the traffic flowing through the network. 
\textcolor{black}{Specifically, it introduces a Day Slice, designed for performance during peak usage hours, and a Night Slice, optimized to reduce power consumption by selectively deactivating router line cards during off-peak times.}
During daytime operations, the slices may follow standard metrics. At night, specific IP cards can be shut down without impacting the night slice. To enable selective IP card operation when defining network slices, the approach may consider the forecast traffic profile and define algorithms for joint optimization of routing and line card activation.

Unlike dynamic adaptation approaches, this work uses predictable daily traffic patterns. Backbone networks show stable differences between day and night traffic. Operators can estimate these patterns from historical data. 

The proposed approach defines two precomputed configurations. One configuration targets peak traffic. The other targets low traffic periods. The network switches between them at predefined times. This reduces the number of reconfigurations. It also limits control-plane changes.

This design avoids continuous routing updates. It reduces instability and operational complexity. It also aligns with current traffic engineering practices in IP/MPLS networks.

The main contributions of this work are:

\begin{itemize}
    \item A time-based IP slicing strategy with two configurations. One slice is used during the day. The other is used at night.
    \item A joint optimization of routing and line-card activation. Link weights are tuned to reduce the number of active cards.
    \item A multi-objective formulation with two metrics. The model considers energy and latency.
    \item An extension to multi-service traffic. The model includes latency constraints for premium traffic.
\end{itemize}

The paper is organized as follows. Section \ref{back} \textcolor{black}{reviews related work on} energy efficiency and network slicing. Section \ref{problem} \textcolor{black}{introduces} the problem and \textcolor{black}{outlines} the proposed solution. Section \ref{experiment} validates the approach in a simplified scenario and a real-world network topology. Finally, Section \ref{conclusion} \textcolor{black}{summarizes} the conclusions.

\section{Background and Related Work}
\label{back}

The rapid evolution of 5G mobile networks has introduced new services that require large computational and communication resources. While 5G improves data rates, latency and connectivity, it also increases the energy consumption of network infrastructures. As operators expand their deployments, improving energy efficiency has become an important objective to reduce operational costs and environmental impact.

Network slicing contributes to this objective by enabling a more efficient use of shared infrastructures. It allows operators to create several logical networks on the same physical platform, each tailored to specific service requirements. Key enablers of slicing include Software Defined Networking (SDN), Virtual Network Functions (VNFs), and cloud-based platforms, which support flexible resource allocation.

Energy efficiency in network slicing can be improved through both hardware and software strategies (\cite{Lorincz24,Tan22}). Hardware-oriented approaches focus on more efficient components and architectures. Software approaches adapt the network configuration to traffic conditions. These strategies include dynamic resource allocation and the selection of device power states.

Dynamic allocation of network resources reduces energy consumption by activating only the elements required to serve current traffic. Idle components can enter low-power states. Sleeping mechanisms extend this concept by allowing network elements to switch to low-power modes when they are not used. These mechanisms can operate at different levels of the infrastructure (\cite{Lorincz24}), including base stations, routers and cloud-based VNFs.

Several works study energy-aware resource allocation for network slicing. In \cite{Masoudi22}, energy consumption is minimized through joint allocation of communication and computation resources across slices. In a similar way, \cite{Matthiesen18} formulates resource allocation in sliced networks as a multi-objective optimisation problem that jointly considers throughput and energy efficiency under QoS constraints.

Energy-aware routing has also received significant attention. Many approaches aim to reduce the number of active network elements by concentrating traffic on a subset of links or nodes. Optimization-based routing algorithms can minimize the set of active devices while meeting capacity and QoS constraints (\cite{chen2018tear}). In SDN environments, centralized control enables routing adjustments that follow traffic variations (\cite{kamboj2022energy}).

Other studies focus on energy-efficient routing in backbone networks. The multi-stage routing framework proposed in \cite{zhang2022multi} reduces the number of active network components while maintaining network performance. Similarly, traffic engineering based greedy routing methods aim to minimize the set of powered-on elements in bundled-link networks (\cite{zhang2024tepg}).

Traffic patterns often follow strong daily variations. Some works adapt routing decisions to these temporal changes. For example, routing adjustment models based on integer programming have been proposed to reduce energy consumption while considering reconfiguration costs (\cite{zhang2020network}). Other algorithms dynamically select energy-efficient paths as flows arrive and depart (\cite{chen2018tear}). Similarly, \cite{Jacob23} evaluates link sleeping and rate adaptation based on daily traffic variations to reduce energy consumption.

Energy savings can also be achieved by deactivating idle hardware during low-demand periods. In IP-over-WDM backbone networks, switching off unused router line cards can significantly reduce energy consumption (\cite{idzikowski2010saving}). Traffic is concentrated on fewer links so that unused components can enter low-power states.

Artificial intelligence techniques have also been studied for energy optimization. Machine learning models can analyse traffic patterns and adjust slice configurations (\cite{Tan22}). Reinforcement learning has been applied to routing decisions that balance energy consumption and performance (\cite{kandil2022qlearning}). Other works investigate SDN deployment strategies that consider link utilisation and network topology (\cite{wang2023energy}).

Despite these advances, most studies address either energy-aware routing or resource allocation in isolation. Few works combine traffic engineering and slicing mechanisms to adapt the network configuration to different traffic periods. The approach proposed in this paper follows this direction. It uses IP-based network slicing and multi-objective optimisation to adapt the network configuration to daily traffic patterns while maintaining service performance.

Daily traffic patterns offer an opportunity to adjust the network configuration over time. During low-demand periods, part of the infrastructure can be deactivated without affecting service quality, while full capacity is required during peak hours. 

The approach proposed in this paper follows this idea. It introduces an IP-based slicing strategy that separates network operation into two temporal configurations: a performance -oriented slice used during peak hours and an energy-efficient slice used during low-demand periods. Multiobjective optimisation is used to balance energy consumption and latency while maintaining the required service performance.

Most existing works focus on dynamic adaptation. They adjust routing or resources as traffic changes. These approaches require frequent updates. They may introduce instability.

This work follows a different approach. It uses a small set of precomputed configurations. Each configuration matches a known traffic period. This reduces the number of changes on the network.

The model also combines routing and resource activation. Many previous works study these problems separately. This work integrates both aspects in a single formulation. It also considers multi-service traffic with latency constraints.




\section{Problem statement and proposed solution}
\label{problem}
A major challenge in 5G network energy efficiency is the fluctuation of network traffic between peak and off-peak hours. The slicing mechanism is IP-based, ensuring that traffic is automatically switched between slices at predefined times, enabling continuous operation while optimizing energy efficiency. Active flows, such as TCP, are maintained during slice switching through flow management techniques to preserve session continuity. Achieving this in operational networks requires careful engineering. In IP/MPLS backbones, reducing disruption when switching between Day and Night slices by changing routing metrics can be done if the new forwarding state is pre-installed in the routers before the old state is removed (“make-before-break”). This can be achieved using techniques such as:
\begin{itemize}
    \item MPLS or Segment Routing with Pre-Provisioned Paths: New slice paths are instantiated and activated before decommissioning the old ones, ensuring packets continue to flow without interruption.
    \item Fast Reroute and BGP PIC: These mechanisms minimize convergence time, reducing the risk of TCP timeouts during the switch.
    \item Flow-Aware Migration: Existing flows may be “pinned” to their current path until completion, while new flows are directed to the new slice.
    \item State Synchronization for Middleboxes: If NAT, firewalls, or DPI are involved, their state must be preserved or replicated across slices to avoid breaking stateful connections.
\end{itemize}

Without such measures, transient packet loss, routing convergence delays, and packet reordering can occur during slice switching, potentially degrading TCP performance or resetting sessions. Therefore, for the approach to be fully seamless in production environments, it is recommended to combine the slice-switching logic with fast-failover and pre-provisioning techniques widely used in carrier networks.

In the scenario under consideration, during the day, mobile networks experience high traffic loads, requiring robust resource allocation to maintain service quality. Conversely, during nighttime hours, network demand significantly decreases, yet traditional networks continue to operate with the same energy consumption settings as during peak hours. This results in unnecessary energy expenditure, leading to higher operational costs and increased environmental impact.

This behavior is consistent with common backbone planning practices, where networks are dimensioned based on peak or busy-hour traffic demand. As a result, the deployed capacity is sufficient for worst-case conditions but remains underutilized during low-demand periods. This gap between provisioned capacity and actual traffic demand motivates the use of energy-aware configurations during off-peak periods.

To address this inefficiency, we propose a dual IP slice approach, where a Day Slice and a Night Slice are dynamically managed to optimize network performance and power consumption based on traffic patterns. The Day Slice \textcolor{black}{assumes shortest path routing for all traffic demands using the default} metrics 
to meet daytime demand. During peak daytime hours, users require consistent connectivity with high data rates. However, maintaining the same resource allocation at night results in energy wastage due to low traffic demand. Thus, the Night Slice is designed to prioritize energy efficiency by reducing power consumption in card-based routers, adapting the routing to the significantly lower traffic volume, and deactivating non-essential resources. In previous studies (\cite{osa24}), the authors proposed and validated a router energy consumption model which shows that switching off line cards for card-based routers reduces power consumption. Thus, the Night Slice will focus on rearranging the metrics and routing paths to reduce the number of active cards. Cards that are non-essential during off-peak hours can be put into low-power or sleep modes, reducing energy consumption without compromising essential connectivity.

\begin{figure}[ht]
    \centering
    \includegraphics[width=0.7\linewidth]{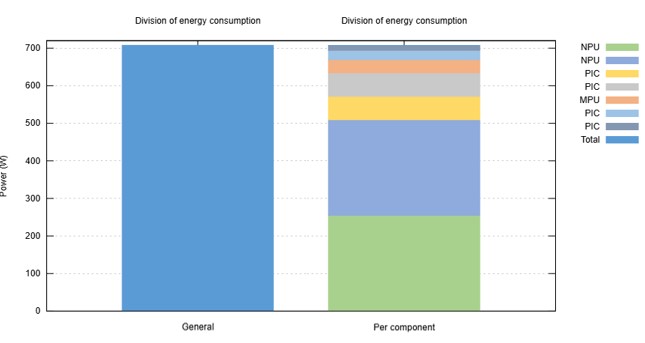}
    \caption{Division of energy consumption by component in a typical router line card.}
    \label{fig:energy_breakdown1}
\end{figure}

\begin{figure}[ht]
    \centering
    \includegraphics[width=200 pt]{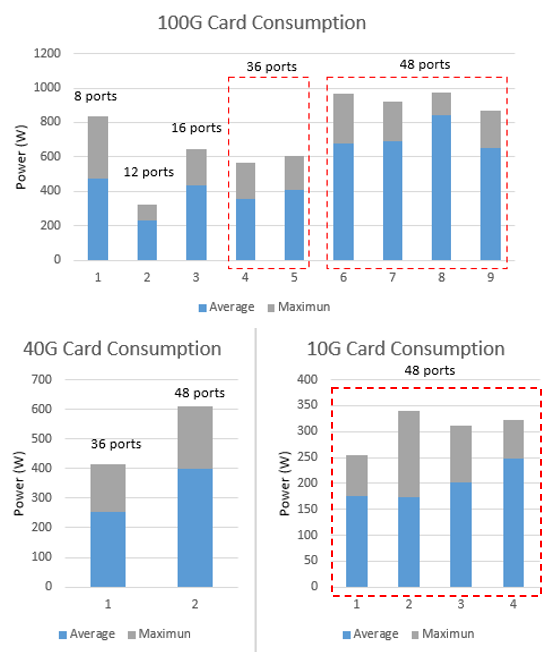}
    \caption{Average and maximum power consumption for 100G, 40G, and 10G cards as a function of the number of ports.}
    \label{fig:energy_breakdown2}
\end{figure}

To further illustrate the potential for energy optimization, Figures  ~\ref{fig:energy_breakdown1} and  ~\ref{fig:energy_breakdown2} together highlight that energy optimization can be achieved both by selectively managing internal card components and by making strategic decisions regarding card types and port densities. This dual approach maximizes energy savings during low-traffic periods while maintaining network performance.

Figure ~\ref{fig:energy_breakdown1} displays how the total energy consumption of a line card is distributed among its internal components, such as the Network Processing Unit (NPU), Physical Interface Card (PIC), and Main Processing Unit (MPU). This breakdown reveals that not all components need to be fully active at all times. As previously pointed out, during low network demand periods, components can be placed in low-power modes to reduce energy consumption without affecting essential connectivity, allowing for dynamic adaptation to traffic conditions. For instance, if only a subset of ports or processing capacity is required during off-peak hours, the corresponding NPUs, PICs, or other subsystems can be powered down or set to low-power states. This strategy complements the overall goal of the Night Slice: to minimize energy usage by both deactivating entire cards when possible and optimizing the operation of active cards at the component level.

Figure~\ref{fig:energy_breakdown2} presents a comparative analysis of the average and maximum power consumption for network cards with different capacities (100G, 40G, and 10G) and varying numbers of ports. These bar charts demonstrate how both the card type and the number of active ports significantly impact total energy usage.

In particular, the results show that 100G cards, despite their higher capacity, are more energy efficient when compared to deploying multiple lower-capacity cards (such as two 40G cards) to achieve a similar overall throughput. For example, a single 100G card with 48 ports consumes less power on average than two 40G cards with 48 ports each, while providing greater or equivalent switching capacity. This efficiency gain is especially relevant for large-scale routers, where consolidating traffic onto fewer high-capacity cards can lead to substantial energy savings.

Furthermore, the charts highlight that as the number of ports increases, the power consumption rises for all card types, but the increase is less pronounced for higher-capacity cards. This observation supports the strategy of optimizing network configurations by favoring high-capacity cards and minimizing the total number of active cards and ports, especially during periods of low traffic demand. Such an approach not only reduces operational costs but also contributes to the sustainability objectives of modern network infrastructures.

The optimization problem aims to minimize the energy consumption of an IP backbone network by selectively deactivating line cards during low-traffic periods, while keeping end-to-end latency and link utilization within acceptable limits.

\subsection{Problem formalisation}
The network is represented as an undirected graph \( G(V, E) \), where \( V \) is the set of nodes (routers) and \( E \) is the set of bidirectional links. 
Each link \( e = (u,v) \in E \) is characterized by a configurable routing weight \( w_e \), physical distance \( d_e \), and associated latency and capacity parameters.

\vspace{0.5em}
\subsubsection{Routing Model}
Traffic between each source–destination pair \( (s,t) \) is routed along the shortest path computed using Dijkstra’s algorithm, where link costs are given by the weights \( w_e \). 
For every demand \( D_{st} \) (in Gbps), the traffic load on each link is accumulated over all shortest paths that traverse it:
\[
T_e = \sum_{(s,t): e \in P_{st}} D_{st}
\]
where \( P_{st} \) is the shortest path between \( s \) and \( t \).

\vspace{0.5em}
\subsubsection{Capacity and Port Model}
Each link has a capacity that depends on the number of active ports and the port rate:
\[
C_e = N_e \cdot R_p,
\]
where \( R_p = 100~\mathrm{Gbps} \) is the port speed and \( N_e \) is the number of active ports on link \( e \). 
The number of ports required is the smallest integer greater than or equal to the ratio of carried traffic to port rate:
\[
N_e = \lceil T_e / R_p \rceil.
\]
Each router line card supports two ports, so the number of cards per link is:
\[
K_e = \lceil N_e / 2 \rceil.
\]
The total number of active cards in the network is then:
\[
M_1(w) = \sum_{e \in E} K_e.
\]

\vspace{0.5em}
\subsubsection{Latency Model}
The average end-to-end latency experienced by traffic is the sum of three components: physical propagation delay, processing delay, and response time.

\begin{itemize}
    \item \textbf{Physical delay (\(D^{phy}_e\))}: proportional to fiber length. Each link (edge $e$) in the SNDlib topology provides a nominal length in kilometers, which is imported as the edge weight in the NetworkX graph.  This physical distance is converted to propagation delay using a fiber latency factor of $0.005~\text{ms/km}$, consistent with the typical speed of light in optical fiber (approximately $2\times10^8~\text{m/s}$). 
    \item \textbf{Processing delay (\(D^{proc}_e\)):} processing headers, route lookup and other router functions, \(0.01~\mathrm{ms}\) per hop.
    \item \textbf{Average response time, including queuing delay (\(D^{que}_e\)):} depends on link utilization and is modeled using a utilization-dependent function inspired by the M/M/1 queue:
    \[
    D^{que}_e = D_{q,0} \cdot \frac{1}{(1 - U_e)^{s_q}},
    \]
    where \( D_{q,0} \) represents a baseline service time (set to 1  $\mu$s for 100 Gbps links), \(U_e\) is the link utilization and \( s_q \ge 1 \) is a sensitivity parameter controlling how sharply delay increases as utilization approaches saturation. 
    For \(s_q = 1\), the expression reduces to the classical M/M/1 model. In this work, we use \(s_q = 1.5\) to better approximate real network behavior, where queuing delay increases more gradually under moderate load while still capturing the sharp rise near high utilization. This formulation provides a computationally efficient and tunable approximation suitable for large-scale optimization.

\end{itemize}

For each path \( P_{st} \), between a source $s$ and a target $t$, the total delay is
\[
L_{st} = \sum_{e \in P_{st}} (D^{phy}_e + D^{proc}_e + D^{que}_e),
\]
and the traffic-weighted average network latency is
\[
M_2(w) = \frac{\sum_{(s,t)} D_{st} \cdot L_{st}}{\sum_{(s,t)} D_{st}}.
\]

\vspace{0.5em}
\subsubsection{Link Utilization}
The utilization of each link is defined as
\[
U_e = \frac{T_e}{C_e},
\]
and the maximum link utilization across the network is
\[
M_3(w) = \max_{e \in E} U_e.
\]

\vspace{0.5em}
\subsubsection{Optimization Objectives and Constraints}

The optimization problem jointly minimizes two primary objectives:
\begin{itemize}
    \item $M_1(w)$ – the total number of active line cards, representing network energy consumption;
    \item $M_2(w)$ – the average end-to-end latency, computed as the traffic-weighted mean delay across all source–destination pairs.
\end{itemize}

Link utilization acts as a feasibility constraint rather than a direct objective. Solutions that saturate any link beyond 85\% of its capacity are discarded. The overall formulation is expressed as:

\[
\begin{aligned}
\min_{w} \quad & \big( M_1(w), \; M_2(w) \big) \\
\text{s.t.} \quad 
& U_e(w) \le 0.85, \quad \forall e \in E, \\
& N_e(w) \le N_e^{day}, \\
& w_e \in [w_{\min}, w_{\max}], \quad \forall e \in E.
\end{aligned}
\]

The second constraint ensures that the number of ports required on each link $N_e(w)$ does not exceed the available capacity in the daytime (reference) configuration $N_e^{day}$. To handle possible violations smoothly, an overcapacity penalty is introduced:

\[
\Pi_{\text{over}} = \sum_{e \in E} \max(0, N_e - N_e^{day}),
\]
which is directly used as an inequality constraint in the optimization. Feasible solutions satisfy $\Pi_{\text{over}} = 0$.

Additionally, when resilience is enabled, the network must remain $k$-edge-connected (typically $k=2$). A connectivity penalty $\Pi_{\text{conn}}$ counts the number of node pairs that fail to meet the required redundancy level. The corresponding constraint is expressed as:
\[
\Pi_{\text{conn}} = 0 \quad \text{(feasible if all pairs satisfy $k$-connectivity).}
\]

The resulting Pareto-optimal front reveals the trade-off between energy savings and performance under these constraints.

\subsection{Algorithms}
Setting up metrics and routes specifically oriented to energy efficiency through line-card deactivation is not a straightforward task, especially in large networks. First, it is possible to look for the maximum savings by searching the solution space in order to minimize the total number of active line cards, which can be quite complex. Metaheuristic algorithms, e.g. differential evolution (DE)\cite{Lampinen2004} or dual annealing (DA)\cite{Xiang97}, are useful when the objective function has multiple local minima (e.g. multiple solutions for minimum active line cards), although they do not guarantee to find the global optimum. DE and DA can be applied to approximate the optimal set of link weights that minimizes the number of active line cards. DE is a population-based optimization algorithm that iteratively improves a population of candidate solutions (e.g. sets of link weights) by combining existing ones according to simple arithmetic rules. DA is a hybrid stochastic optimization method that combines a broad, temperature-driven global search with focused local refinement.

To address the multi-objective optimization, more complex algorithms are required.

Because the objective space is nonlinear and nonconvex, gradient-based solvers are inadequate. 

Three state-of-the-art evolutionary algorithms were selected for comparison due to their complementary strengths in multi-objective optimization. 
NSGA-II\cite{NSGAII} (Non-dominated Sorting Genetic Algorithm II) employs elitist non-dominated sorting and crowding distance to preserve diversity along the Pareto front. 
CTAEA\cite{CTAEA} (Constrained Two-Archive Evolutionary Algorithm) maintains two archives: one for convergence and one for diversity. This separation helps handle constraints and keeps a balance between finding good solutions and exploring the search space. 
AGE-MOEA\cite{AGEMOEA} (Adapted Geometry Estimation based Multi-Objective Evolutionary Algorithm) uses geometry estimation of the Pareto front to guide convergence efficiently. 
These methods were chosen to represent distinct approaches to Pareto dominance, constraint management, and diversity preservation.

Each individual represents a vector of link weights. 
For each generation, network metrics are evaluated using a parallelized Python implementation based on \texttt{networkx} and \texttt{numpy}.
All experiments were run with population sizes of 150 and up to 80 generations.

In the following section, different scenarios and solutions are analyzed.

\section{Experimental evaluation}
\label{experiment}
\subsection{Maximum savings}
In this first experiment, we approximate the maximum savings in terms of line cards that can be set to inactive or idle without taking into account other requirements such as latency.

\begin{figure}[htbp]
    \centering
    \includegraphics [width=0.6\textwidth, trim=0.5cm 1.3cm 0.5cm 1.3cm, clip]{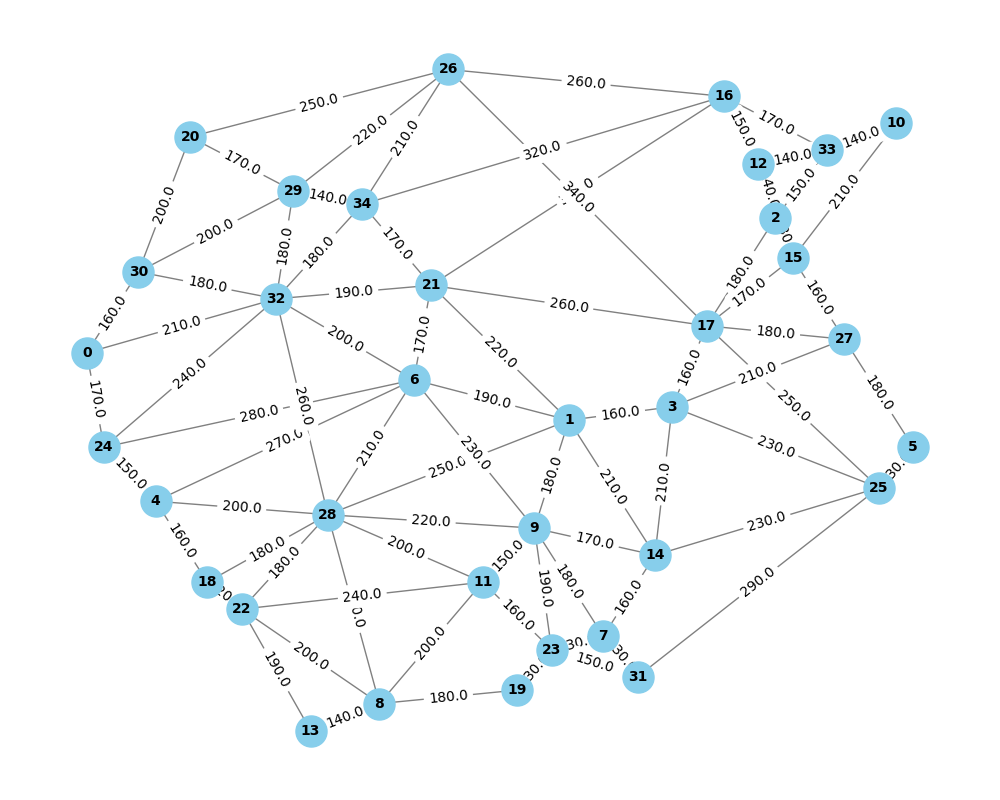}
    \caption{India35 topology from SNDLib\cite{sndlib10}}
    \label{fig:india35}
\end{figure}

We select the \texttt{india35} topology (see Figure \ref{fig:india35}) from the Survivable fixed telecommunication Network Design library (SNDlib) \cite{sndlib10}. This topology represents a medium-scale IP backbone with realistic geographic distances and heterogeneous connectivity.

The reference traffic matrix provided in SNDlib is used as the demand model. Although this matrix is not explicitly defined as peak demand, we interpret it as representative of dimensioning conditions. This assumption is consistent with common backbone planning practices, where networks are provisioned to handle worst-case or near-peak traffic levels.

For the night traffic profile, we simulate reduced demand by scaling the reference matrix. This allows us to evaluate how much of the provisioned capacity can be temporarily deactivated during low-demand periods. The scaling factor ranges from 0\% to 20\%.

For simplicity, we considered symmetric bidirectional traffic and symmetric routing (same traffic in both link directions). Shortest paths are identified for each traffic demand using Dijkstra algorithm. Then, based on the demands and the paths selected, the accumulated traffic per link is calculated. Based on this information, and considering 100 Gbps ports, the required number of ports at each end of the link to process the traffic is computed.

Once we know the total number of active ports per node per link, it is possible to accumulate the number of ports required per node and the corresponding active cards. 

For the night traffic profile, we simulated a reduction by multiplying the daytime traffic by a coefficient between 0 and 0.2, thereby scaling the traffic down to 0–20\% of its original level. 
Three different approaches were run. First, the reference scenario with the day metrics is run to identify the active cards without optimization. After this, we adjusted the default routing metrics to steer traffic in a way that allows more line cards to be turned off. 

Differential evolution (DE) and dual annealing (DA) were used to find a set of metric values that minimize the number of active cards (using the implementations of the Scipy library). 

DE algorithm maintains a population of candidate solutions (sets of weights), which evolve over several generations through three steps: mutation, recombination, and selection. First, a population (pool) of different random sets of link weights ($x_1 ... x_p$) is created. The total number of elements $p$ in this first population is calculated as $p=popsize*N$, where $popsize$ is a parameter of the algorithm and $N$ is the number of links of the network. In our first experiments we kept the default $popsize=15$. In each iteration, the best solution $x_{best}$ within the population that minimizes the number of line cards is selected. Then, for each element $x_i$ in the population, a new evolved element is created following the three commented steps. The implementation utilized in this paper (default one for the library used) uses a default $best1bin$ mutation strategy, where $x_{best}$ is combined with the weighted difference of two other randomly chosen individuals as in $mv_i = x_{best} + F * (x_{r1} - x_{r2})$, where $mv_i$ is the mutated vector, $x_{r1}$ and $x_{r2}$ are the two random sets of weights (with $r1$ and $r2$ different from $i$) and $F$ is the mutation factor (where we set a value of 0.5).

A recombination rate, $0.5$ in our experiment, controls how much of the mutated vector $mv_i$ replaces the original candidate solution $x_i$, i.e. how many of the link weights (in average) from the element are replaced by those of the mutated vector, in this case 50\% probability for each weight to be replaced. This generates a new trial candidate $xt_i$ from each original one $x_i$

After that, the selection step compares the results when applying the trial $xt_i$ set of weights and the original $x_i$: the better one survives into the population of the next generation.

The algorithm iterates until the standard deviation of the objective values in the current population is less than a threshold, or until the maximum number of generations is reached. ($maxiter=1000$ is the default value).

DA behaves in a different way that DE. DA is a trajectory-based optimizer, not a population-based one, so it maintains exactly one solution (set of weights) at a time. The algorithm starts by randomly selecting a candidate solution. In the global exploration phase, a new trial point is generated by adding a random perturbation to the previous solution. The initial $temperature$ sets the scale of the perturbations, with higher temperatures producing larger, more exploratory jumps. The $visit$ parameter controls how likely the algorithm is to attempt occasional long‐distance moves versus smaller, local variations. The $temperature$ decreases through an internal cooling process at each iteration, but the $visit$ parameter remains constant. If the new candidate is better, the algorithm replaces the original, but if it is worse, there is still a probability of acceptance that decreases with each iterations. The $acceptance$ parameter defines the likelihood of making a move that worsens the result so more negative values make uphill moves rarer. A local refinement is then executed, where a standard local optimization method (by default, L-BFGS-B) fine-tunes the best solution found so far.
The algorithm stops when a convergence criterion is met, a maximum number of iterations ($maxiter$) is reached, or the total function evaluations exceed a parameter ($maxfun$).
We used the default parameters from the library, where DA uses an initial temperature of 5230.0, a visit parameter of 2.62, and an acceptance parameter of -5.0.

\begin{figure}[htbp]
    \centering
    \includegraphics [width=0.30\textwidth, trim=0.5cm 1.3cm 1cm 1.3cm, clip]{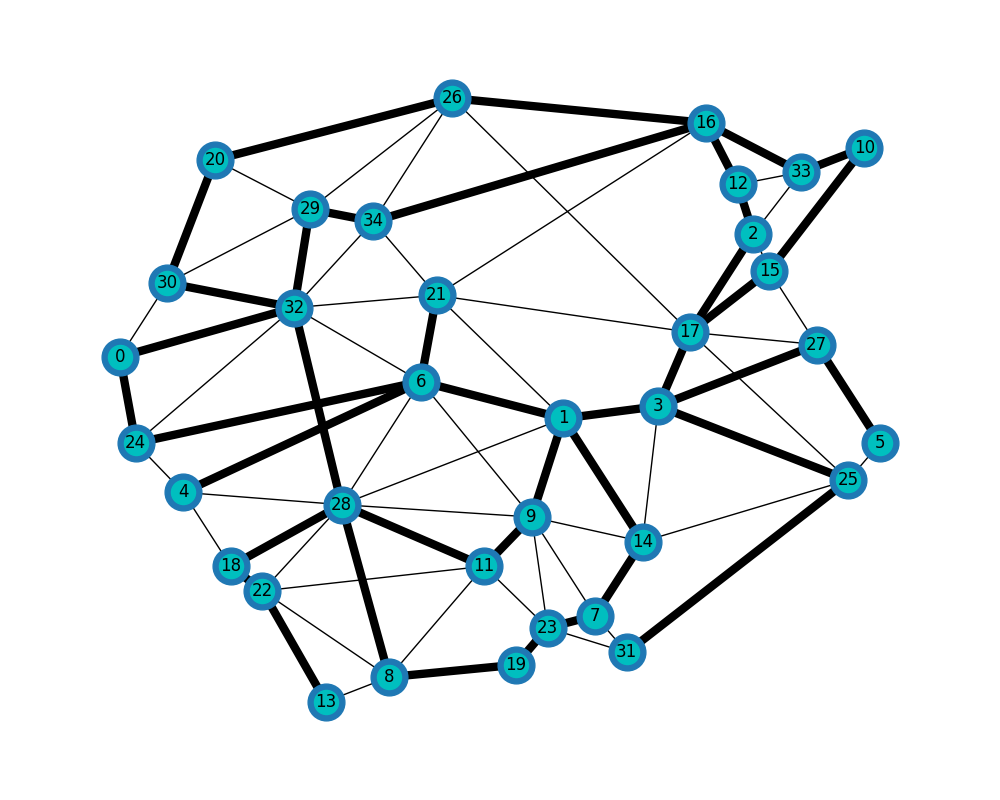}
    \caption{Links with active ports for the DE solution}
    \label{fig:india35_DE}
\end{figure}

\begin{figure}[htbp]
    \centering
    \includegraphics [width=0.30\textwidth, trim=0.5cm 1.3cm 1cm 1.3cm, clip]{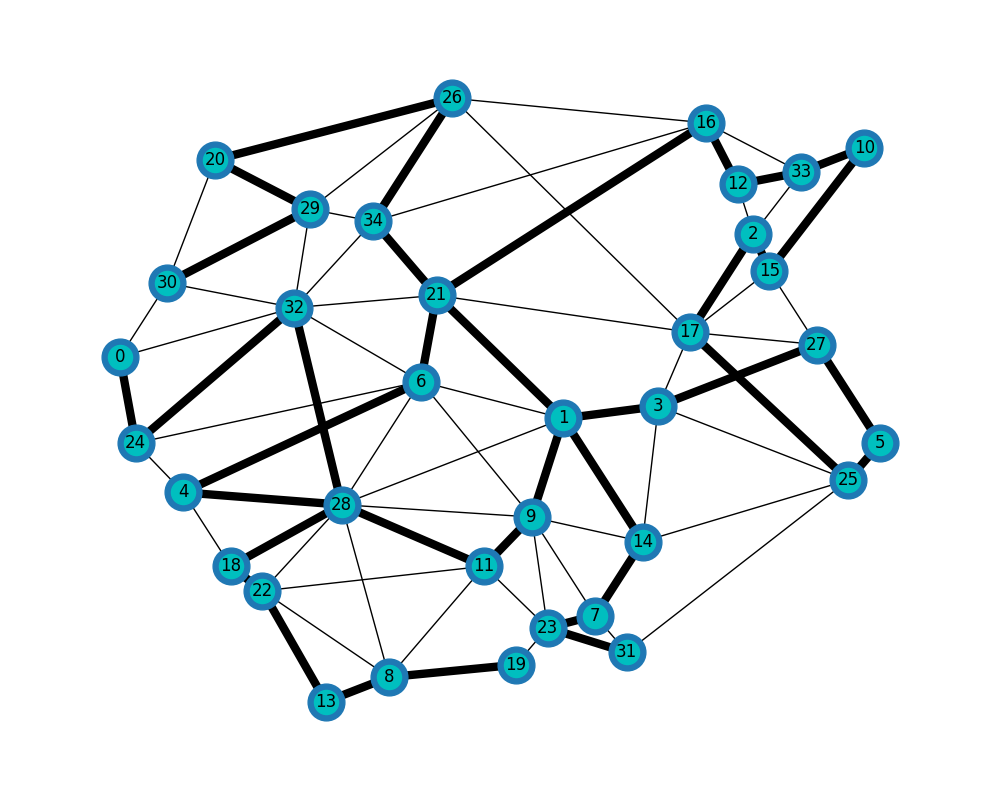}
    \caption{Links with active ports for the DA solution}
    \label{fig:india35_DA}
\end{figure}

The results are shown in Table \ref{tab:india_resources} where the first column provides the resources required by the Day Slice, the second column identifies the number of active ports and cards used when keeping the same metric in the Night Slice, and the third and fourth column, the results of the DE and DA algorithms. The DE algorithm reduces the number of active ports in a 51.3\% and the active cards in 53.3\% while the DA algorithm finds a solution with a 55\% and 56.7\% reduction in active ports and cards respectively. Figures \ref{fig:india35_DE} and \ref{fig:india35_DA} show the links with active ports in the DE and DA solutions respectively. 

\begin{figure}[htbp]
    \centering
    \includegraphics [width=0.30\textwidth, trim=0.5cm 1.3cm 1cm 1.3cm, clip]{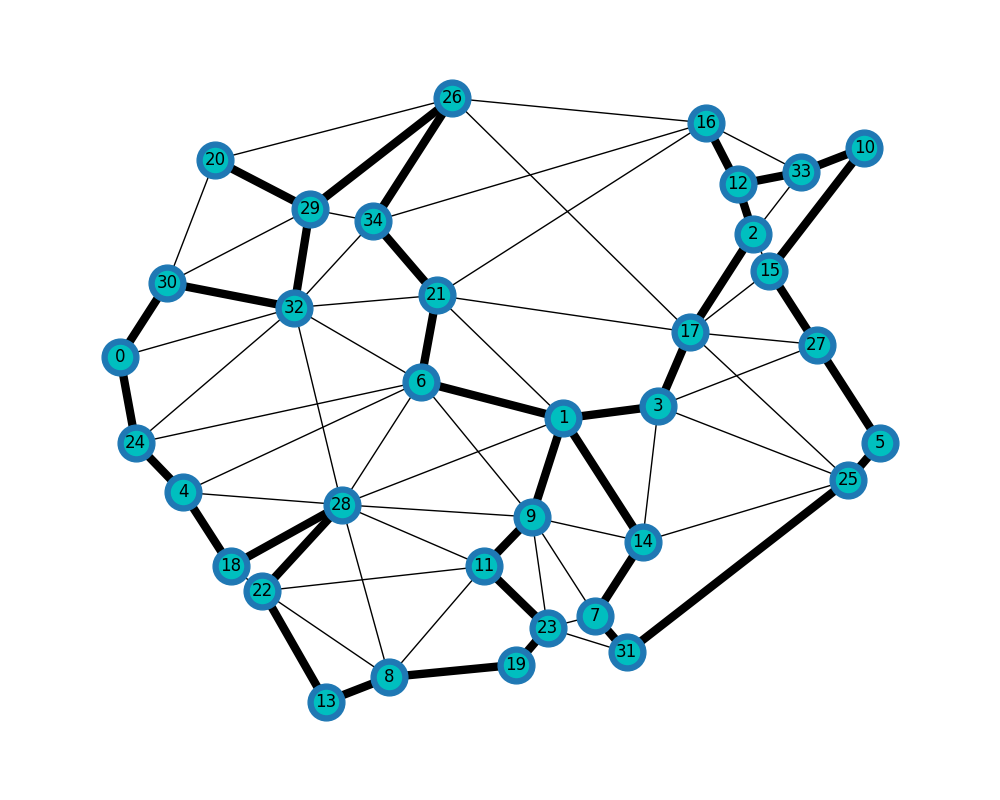}
    \caption{Path automatically generated by Networkx}
    \label{fig:india35_path}
\end{figure}

As DE and DA can be slow to converge and finding the appropriate hyperparameters may be difficult, we looked for a faster solution that could provide similar results. Thus, the last column corresponds to this approach, based on the observation that keeping a single path through the set of nodes minimizes the required cards.
We want to find one single route, let's call it active path, that visits every node in the topology, ideally avoiding visiting each node twice (Hamiltonian path). Due to the computational complexity of the problem and the possibility of such path not existing, an approximation function\footnote{$networkx.approximation.traveling\_salesman\_problem$} from the Networkx library is used which results in a single end-to-end path covering every node with minimal duplication.
Figure \ref{fig:india35_path} shows the path found by the approximate algorithm. Notice that some modifications can be done to visit each node just once, such as changing the 26-29 link to 26-20, the 12-33 to 16-33 and the 9-1-14 to 9-14.\textcolor{black}{ However to maintain the automation of the experiment, the original path was kept}. Low metric values are set for the links of the generated path while large values are assigned to the rest of the links. With the originally generated path, the savings in terms of active cards is 52.2\%. We also confirm that the traffic in every link requires less resources than the ones used in the Day Slice avoiding installing additional cards. In all scenarios, the number of required ports per link remains within the limits used during the Day Slice. 

\begin{table}[htbp]
\caption{Total resources required}
\begin{center}
\begin{tabular}{|c|c|c|c|c|c|}
\hline
\textbf{Resource} & \textbf{Day Slice}& \textbf{Night}
& \textbf{DE} & \textbf{DA} & \textbf{Path}\\
\hline
Active Ports & 278 & 160 & 78 & 72 & 80 \\
\hline
Active Cards & 151 & 90 & 42 & 38 & 43 \\
\hline
\end{tabular}
\label{tab:india_resources}
\end{center}
\end{table}

\vspace{-0.8cm}


\begin{figure}[htbp]
    \centering
    \includegraphics [width=0.48\textwidth]{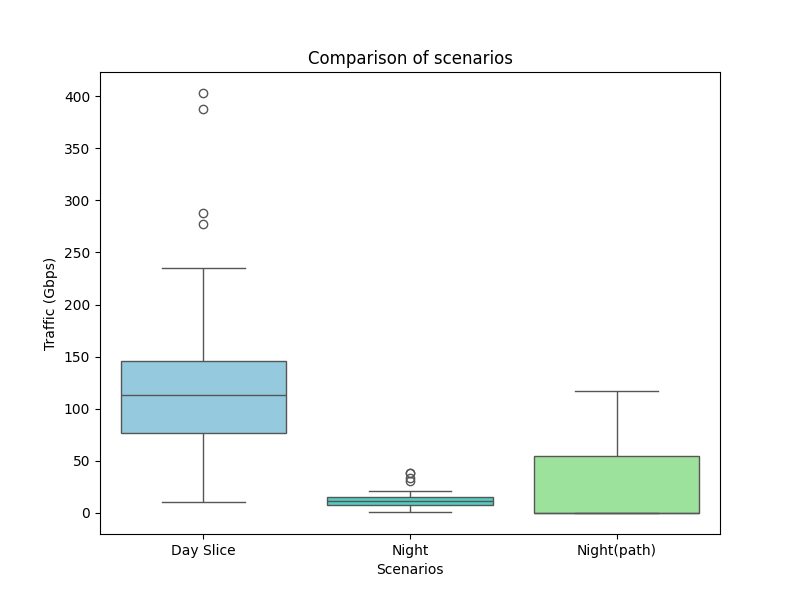}
    \caption{Traffic per link in different scenarios}
    \label{fig:traffic_box}
\end{figure}

\vspace{0.8cm}
The reason for these savings is the very low traffic through the different links at night, as shown in Figure \ref{fig:traffic_box}. The first boxplot shows the distribution of traffic during the day in the different links in the topology. The second one shows the values for the night traffic when the original metrics are used. Finally, the third one shows the traffic when the automatically generated path is used.

\subsection{Multi-objective optimization}
The multi-objective optimization experiments also use the \texttt{india35} topology. 
Its moderate size (35 nodes, 80 links) allows exhaustive traffic simulation and Pareto-based optimization without excessive computational cost.
 
Same configuration is used in terms of capacity of the ports, line card configurations and daytime and nighttime traffic demands. Latency was modeled as the sum of fiber propagation delay (0.005~ms/km), processing delay (0.01~ms per hop), and a utilization-dependent queuing delay. The latter is computed using a parametrized formulation based on link utilization, with a baseline delay of 0.001 ms and a sensitivity exponent of 1.5, as described in Section \ref{problem}. 

The baseline (unoptimized) Night Slice required 90 active line cards, with an average latency of 2.94~ms and maximum utilization of 38.3\%.

\begin{figure}[t]
    \centering
    \includegraphics[width=0.8\linewidth]{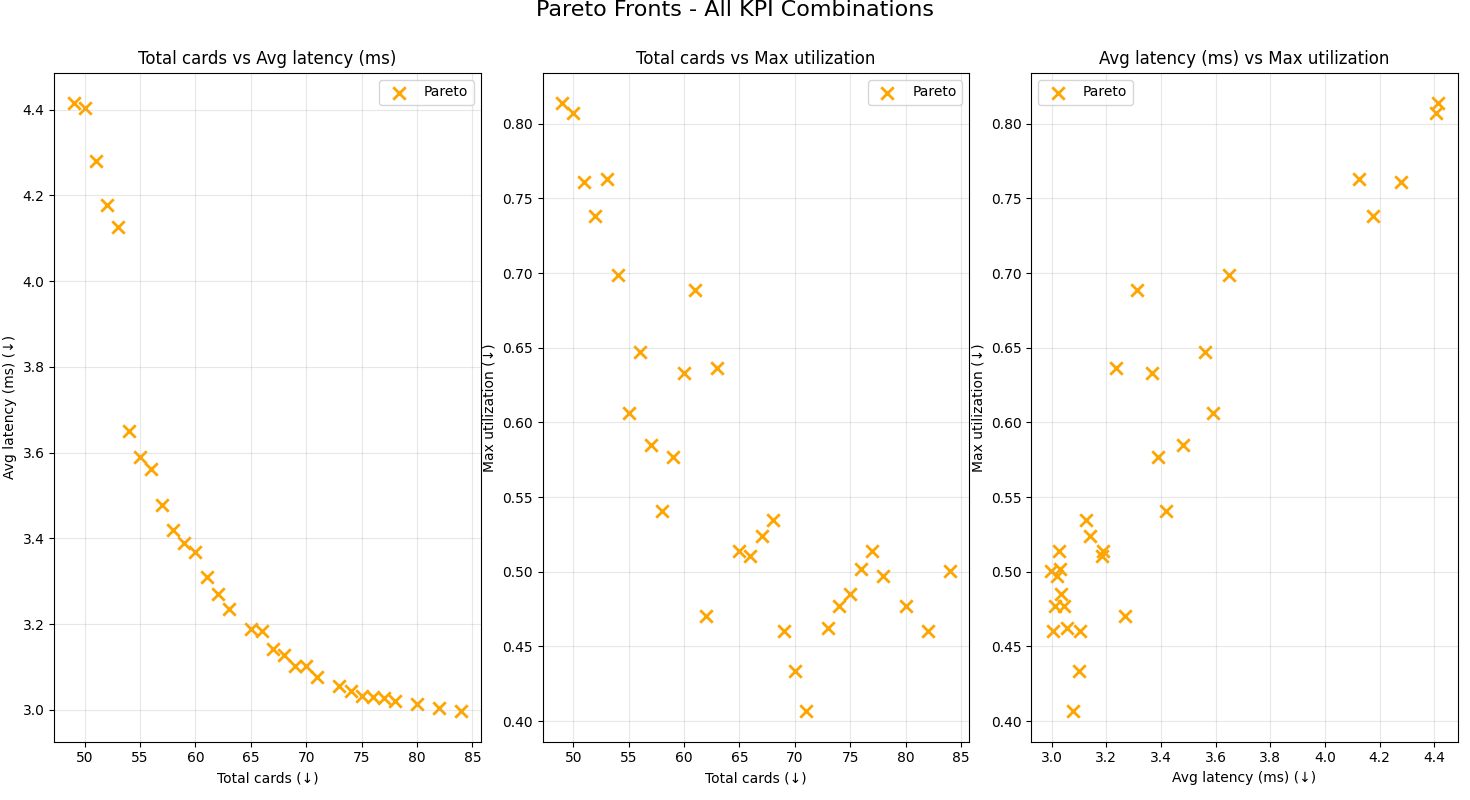}
    \caption{Pareto front obtained with the NSGA-II algorithm.}
    \label{fig:NSGA2_Pareto}
\end{figure}

\begin{figure}[t]
    \centering
    \includegraphics[width=0.8\linewidth]{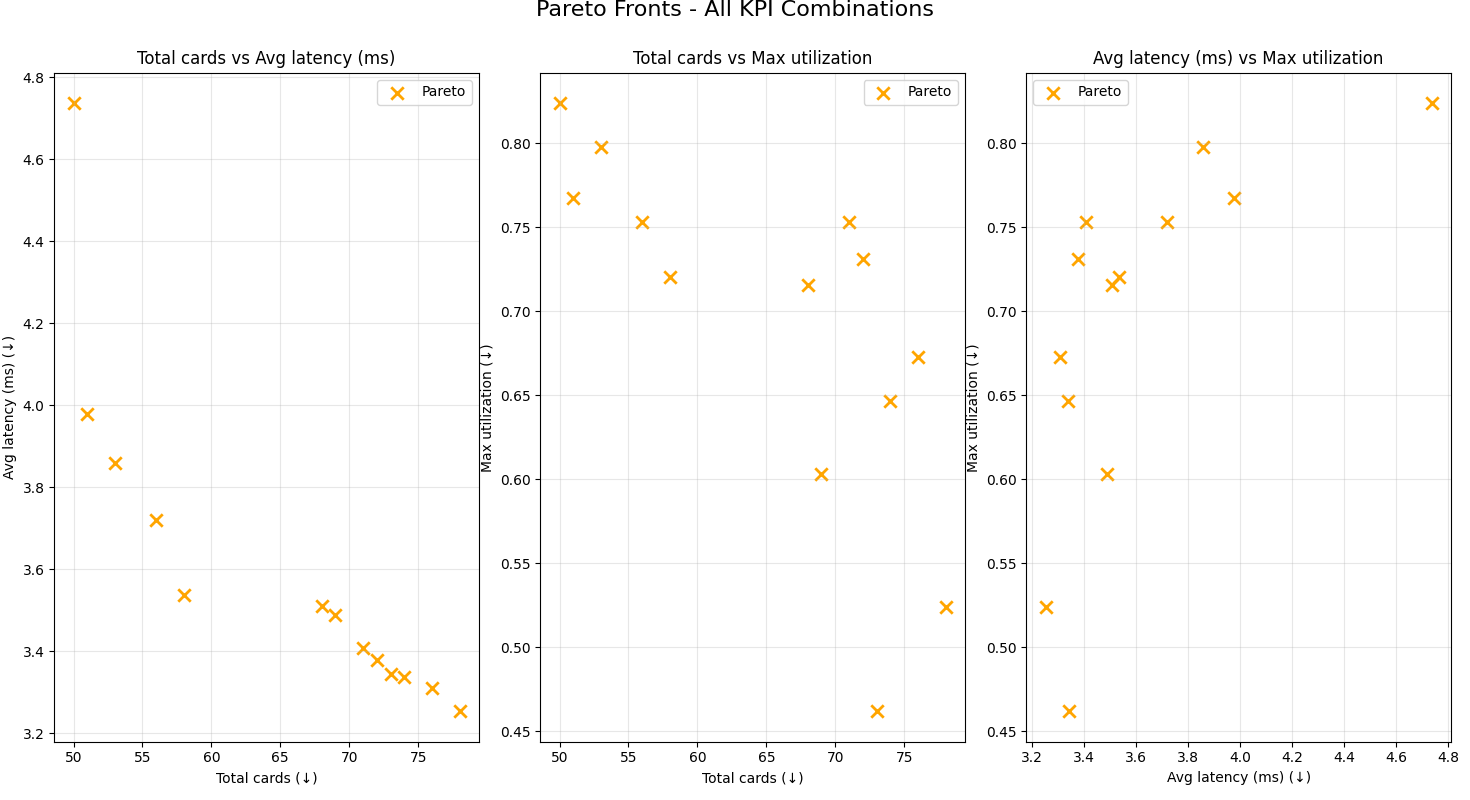}
    \caption{Pareto front obtained with the CTAEA algorithm.}
    \label{fig:CTAEA_Pareto}
\end{figure}

\begin{figure}[t]
    \centering
    \includegraphics[width=0.8\linewidth]{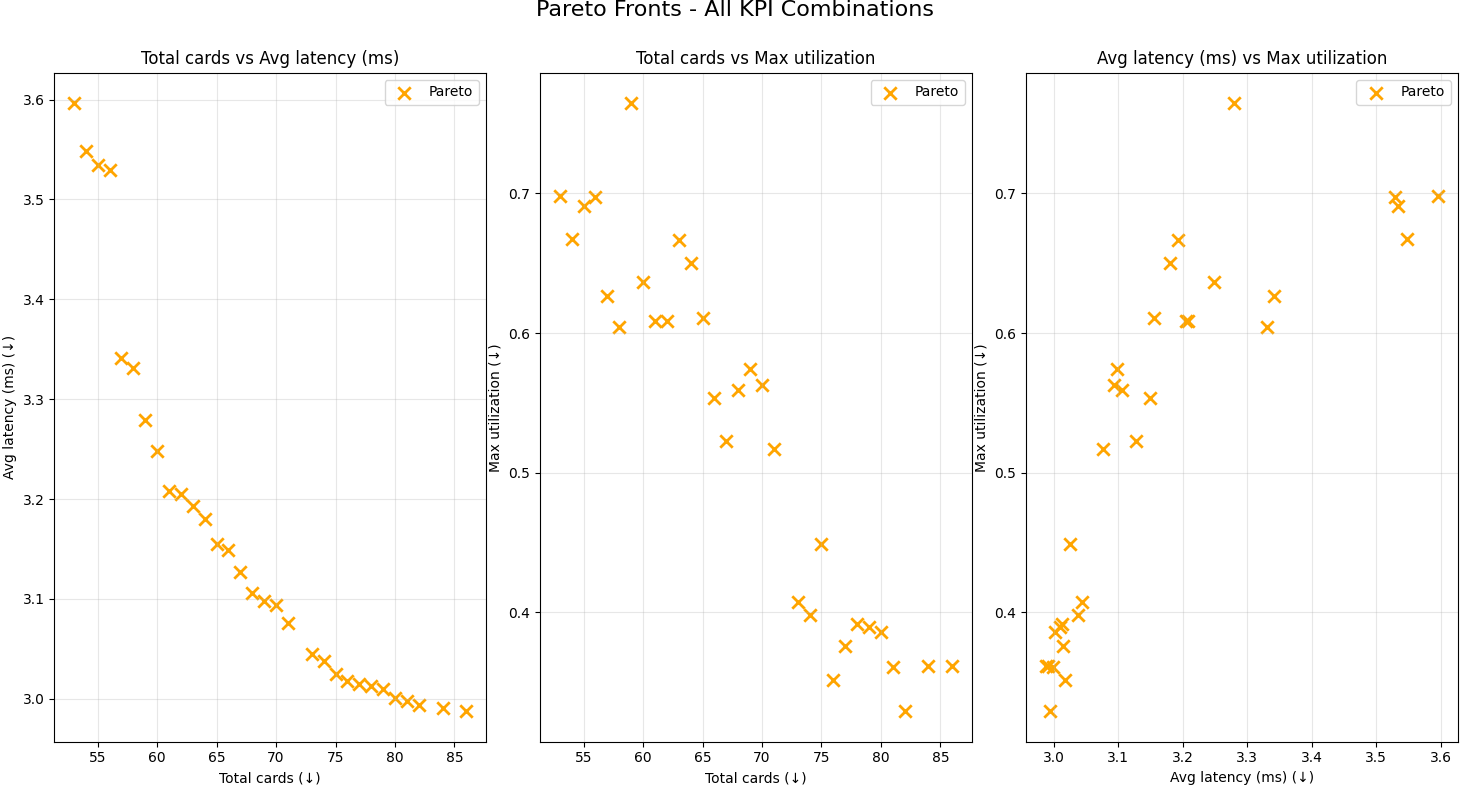}
    \caption{Pareto front obtained with the AGE-MOEA algorithm.}
    \label{fig:AGE_Pareto}
\end{figure}

NSGA-II generated 31 non-dominated solutions after filtering. The results are shown in Figure \ref{fig:NSGA2_Pareto}. CTAEA produced 13 Pareto-optimal solutions (Figure \ref{fig:CTAEA_Pareto}) while AGE-MOEA also generated 31 solutions on the Pareto front (Figure \ref{fig:AGE_Pareto}).

Table~\ref{tab:compare} compares the main outcomes across algorithms.

\begin{table}[ht!]
\centering
\caption{Comparison of Multi-objective Algorithms for India35}
\label{tab:compare}
\begin{tabular}{lccc}
\toprule
Algorithm & Best (Cards) & Best (Latency) \\
\midrule
NSGA-II & 49 (4.41~ms) & 84 (3.00~ms) \\
CTAEA & 50 (4.74~ms) & 78 (3.25~ms) \\
AGE-MOEA & 53 (3.60~ms) & 86 (2.99~ms) \\
\bottomrule
\end{tabular}
\end{table}

NSGA-II found the most energy-efficient configuration (49 cards), but at the expense of higher latency. This means that up to 45\% of router cards can be powered down at night when selecting the best approach in terms of cards. 
NSGA-II and AGE-MOEA provided smoother Pareto fronts with more balanced performance. 

\begin{figure}[t]
    \centering
    \includegraphics[width=0.8\linewidth]{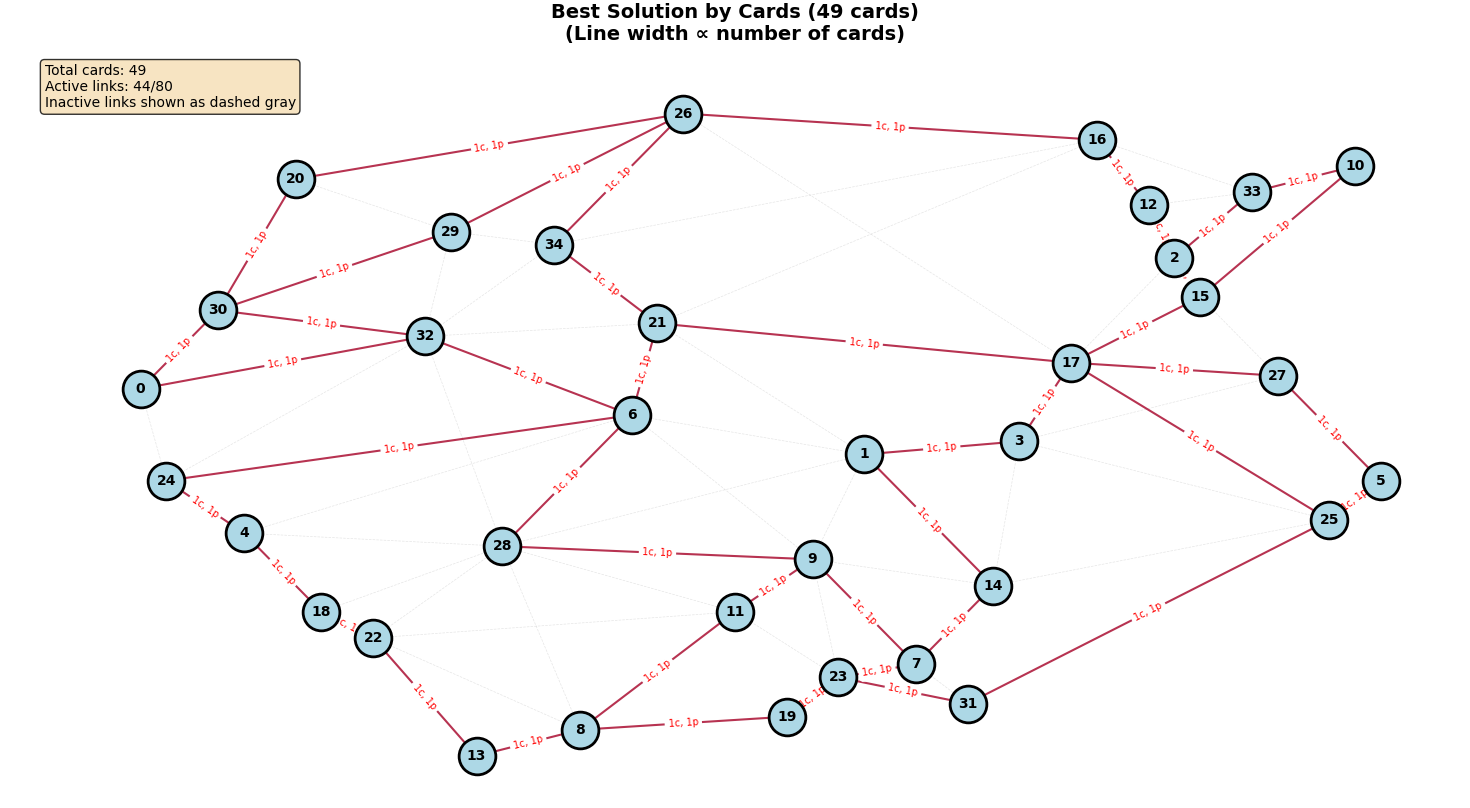}
    \caption{Active paths for the best solution in terms of cards.}
    \label{fig:best_cards}
\end{figure}

Figure \ref{fig:best_cards} show the paths that are active (in red) for the solution with fewer number of active cards (49). 

The next step was to increase the day traffic by multiplying the demand included in the SNDlib file by different factors to understand how it affects the optimization of this topology. We assume that additional ports and cards can be added when required without underlying physical capacity limitation. 

We decided to group in this case the results for all the algorithms for simplicity in Table \ref{tab:result_factor}, comparing to the baseline. The table shows that the algorithms continue finding different solutions to save cards at a cost of increasing the latency. It is possible to look for a balance between both parameters in all cases. Notice that, when the maximum utilization of the links in the baseline nighttime approach increases, the savings in terms of cards decreases. The reduction appears to be non-linear and accelerates as the utilization percentage increases. This makes sense as a higher utilization in the baseline means that the number of cards are near the optimum value.

\begin{table}[!t]
\centering
\caption{Results when increasing traffic (Night Slice)}
\label{tab:result_factor}
\begin{tabular}{lcccc}
\toprule
\textbf{Factor} & \textbf{Type} &\textbf{Cards} & \textbf{Latency (ms)} & \textbf{Max Util.} \\
\midrule
\multirow{2}{*}{x1} & Baseline & 90 & 2.94 & 38.3 \% \\
& Algorithm  & 49--86 & 2.99--4.74 & 33--82\%  \\
\midrule
\multirow{2}{*}{x2} & Baseline & 90 & 2.94 & 76.6 \% \\
& Algorithm  & 60--90 & 3.03-3.73 & 71--83.9\% \\
\midrule
\multirow{2}{*}{x3} & Baseline & 92 & 3.21 & 99.5 \% \\
& Algorithm  & 75--88 & 3.07-3.29 & 75.9--84.9\% \\
\midrule
\multirow{2}{*}{x4} & Baseline & 95 & 2.95 & 88.0 \% \\
& Algorithm  & 85--91 & 3.43-3.63 & 84.1--84.8\% \\
\bottomrule
\end{tabular}
\end{table}

We repeated the original experiment with 3 other topologies from SNDLib: Polska, Germany50 and Nobel-US.

Table \ref{tab:singleservice_results} shows the ranges of values for the different algorithms and compares them to the baseline (using predefined weights). For most topologies, the algorithms identify several solutions that offer different trade-offs between the number of active cards and latency. NSGA-II finds the best solution in terms of cards, except for the Nobel-US topology, where CTAEA finds a better result (and also a slightly wider number of solutions in the pareto front).


\begin{table}[!t]
\centering
\caption{Summary of Single-Service Optimization Results (Night Slice)}
\label{tab:singleservice_results}
\begin{tabular}{lcccc}
\toprule
\textbf{Algorithm} & \textbf{Cards} & \textbf{Latency (ms)} & \textbf{Max Util.} & \textbf{$\Delta$Energy} \\
\midrule
\multicolumn{5}{l}{\textbf{India35}}\\
\midrule
Baseline & 90 & 2.94 & 0.38 & 0 \% \\
NSGA-II  & 49--84 & 3--4.41 & 0.41--0.81 & --46\% (best) \\
CTAEA    & 50--78 & 3.25--4.74 & 0.46--0.82 & --44\% (best) \\
AGE-MOEA & 53--86 & 2.99--3.60 & 0.33--0.77 & --42\% (best) \\
\toprule
{\textbf{Polska}}\\
\midrule
Baseline & 33 & 2.69 & 0.973 & 0\% \\
NSGA-II  & 31--37 & 2.68--2.83 & 0.81--0.85 & --6.1\% (best) \\
CTAEA    & 32--36 & 2.70-2.88 & 0.82--0.85 & --3.0\% (best) \\
AGE-MOEA & 31--37 & 2.68--2.83 & 0.79--0.85 & --6.1\% (best) \\
\toprule
\multicolumn{5}{l}{\textbf{Germany50}}\\
\midrule
Baseline & 101 & 1.71 & 0.21 & 0\% \\
NSGA-II  & 64--97 & 1.64--2.38 & 0.26--0.49 & --37\% (best) \\
CTAEA    & 68--92 & 1.73--2.60 & 0.29--0.55 & --33\% (best) \\
AGE-MOEA & 71--98 & 1.62--1.86 & 0.27--0.47 & --30\% (best) \\
\toprule
\multicolumn{5}{l}{\textbf{Nobel-US}}\\
\midrule
Baseline & 26 & 80.90 & 0.98 & 0\%\\
NSGA-II  & 22--28 & 81.10--82.20 & 0.83--0.84 & --15.4\% (best) \\
CTAEA    & 20--26 & 82.22--91.68 & 0.82--0.85 & --23.1\% (best) \\
AGE-MOEA & 22--27 & 81.00--82.15 & 0.84--0.84 & --15.4\% (best) \\
\bottomrule
\end{tabular}
\end{table}

From a practical perspective, these results suggest that multi-objective link-weight optimization could be integrated into the traffic engineering workflow of large ISPs. 

Network operators could precompute Pareto-optimal configurations for different traffic periods (e.g., day vs. night) and switch between them dynamically. 
However, implementing such adaptive slicing requires coordination with routing protocols (e.g., OSPF or IS-IS) and mechanisms to avoid transient routing loops during reconfiguration.

These experiments focused on a single-service optimization, where all traffic demands were aggregated into a single class.  
Although these results confirmed substantial energy savings,
we focused on average latency, while in real services we might want to ensure that a maximum latency is not reached for some classes of services. This motivated a new formulation for
services with different quality requirements.

\subsection{Multi-Service Optimization with Latency Constraints}

To incorporate differentiated Quality of Service (QoS), the
optimization problem was extended to handle two service
classes with distinct performance targets:
\begin{itemize}
    \item \textbf{Premium service (15\% of total traffic)}:
    delay-sensitive traffic, where we might want to optimize the maximum latency or set up a hard constraint on this value. 
    \item \textbf{Standard service (85\% of total traffic)}:
    best-effort traffic without explicit delay limits.
\end{itemize}

Each service maintains its own routing weights but shares
the same physical topology and port capacities.  The per-service
traffic matrices are obtained by splitting each
source--destination demand~$D_{st}$ according to the normalized
weights $w_s$ of the service mix:
\[
D^{(s)}_{st} = w_s \, D_{st},
\quad \text{with} \quad
w_\mathrm{premium}=0.15,~w_\mathrm{standard}=0.85.
\]
The optimization therefore captures realistic multi-slice
behavior, where premium traffic follows more latency-sensitive
routes while standard traffic absorbs the residual load.


We use again NSGA-II, AGE-MOEA and CTAEA. Two objectives were optimized: energy (active cards) and average premium latency.
The main parameters were:
\begin{itemize}
    \item Port rate: 100~Gbps; 2~ports per card.
    \item Generations: 100.
    \item Constraints: $U_e \le 0.85$, $k$-edge connectivity ($k=2$).
\end{itemize}

Additionally, NSGA-II and AGE-MOEA use a population size of 400. CTAEA uses reference directions generated with the Das–Dennis method. 

We first consider a baseline case without any constraint on the maximum latency of the premium service. This allows the algorithms to explore the trade-off between energy consumption and delay.

\begin{figure}[ht]
    \centering
    \includegraphics[width=250 pt]{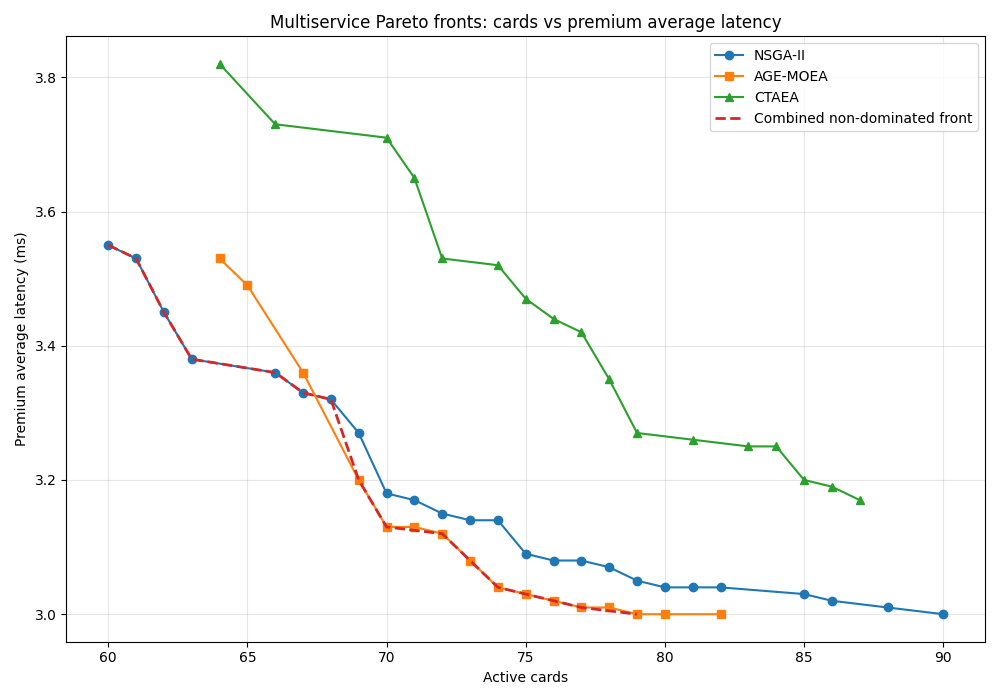}
    \caption{Pareto fronts (cards vs average premium latency) without latency constraint.}
    \label{fig:front_avg}
\end{figure}

\begin{figure}[ht]
    \centering
    \includegraphics[width=250 pt]{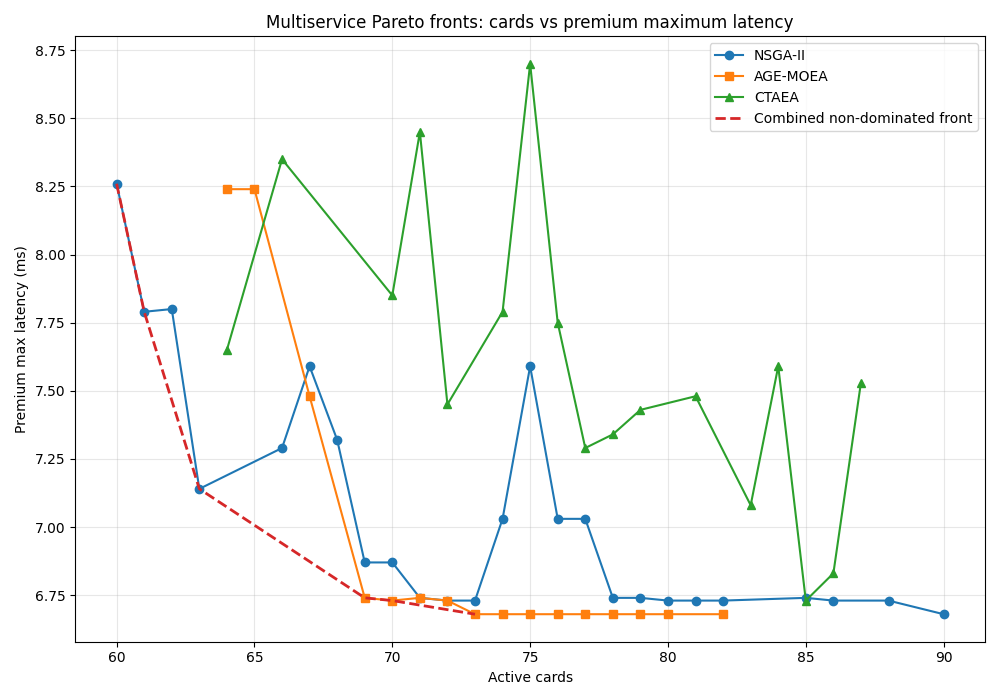}
    \caption{Pareto fronts (cards vs max premium latency) without latency constraint.}
    \label{fig:front_max}
\end{figure}

Figures \ref{fig:front_avg} and \ref{fig:front_max} show the Pareto fronts obtained in this case, using average and maximum premium latency, respectively. Both figures show a clear trade-off between the number of active cards and latency.
Figure \ref{fig:front_avg} shows that NSGA-II explores the widest range of solutions. It finds configurations with the lowest number of cards, down to around 60. These solutions have higher average latency. AGE-MOEA produces a more compact set of solutions. In the range of 64 to 76 cards, it achieves lower average latency and offers a better balance between energy and performance.

Figure \ref{fig:front_max} highlights the impact on maximum latency. The difference between algorithms becomes more visible in this case. Solutions with a low number of cards often lead to higher maximum delays. AGE-MOEA maintains lower maximum latency in the central region of the Pareto front. NSGA-II includes more extreme solutions, but with worse worst-case delay.
CTAEA depends strongly on its configuration. Even with a large number of reference directions, it does not approximate the Pareto front well. Its solutions are more scattered and are mostly dominated by those from NSGA-II and AGE-MOEA. It does not contribute additional non-dominated points in either figure.
These results show that NSGA-II is effective for exploring low-card configurations, while AGE-MOEA provides more balanced solutions, especially when maximum latency is considered.

In practice, networks often require strict delay guarantees for premium traffic. We now evaluate the impact of enforcing a strict latency constraint on the premium service. The maximum premium latency is limited to 7 ms. Only solutions satisfying this constraint are considered feasible. This reduces the feasible solution space and changes the trade-off.

\begin{figure}[ht]
    \centering
    \includegraphics[width=250 pt]{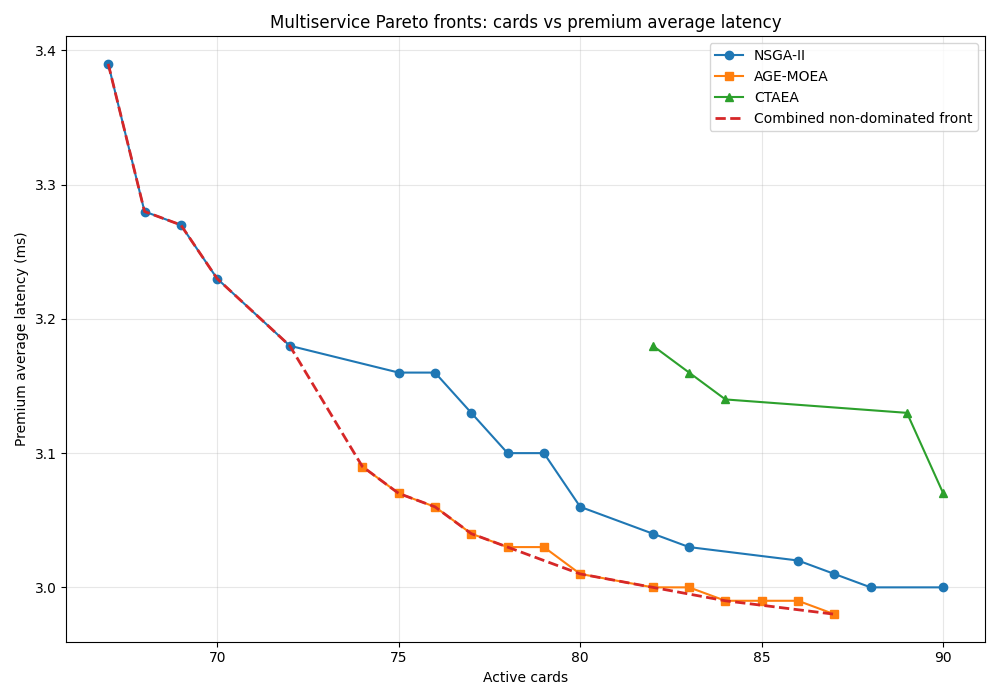}
    \caption{Pareto fronts (cards vs average premium latency) with latency constraint.}
    \label{fig:front_avg_restrict}
\end{figure}

\begin{figure}[ht]
    \centering
    \includegraphics[width=250 pt]{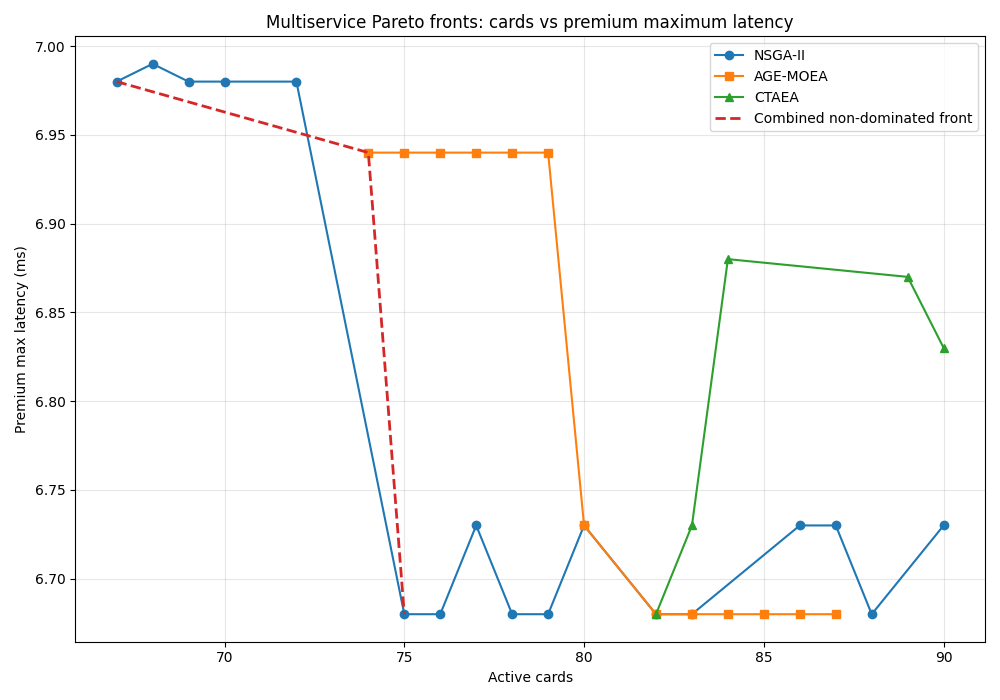}
    \caption{Pareto fronts (cards vs max premium latency) without latency constraint.}
    \label{fig:front_max_restrict}
\end{figure}

Figures \ref{fig:front_avg_restrict} and \ref{fig:front_max_restrict}  show the resulting Pareto fronts. Figure \ref{fig:front_avg_restrict} presents the number of active cards versus the average premium latency. Figure \ref{fig:front_max_restrict} shows the same solutions using maximum premium latency.

In Figure \ref{fig:front_avg_restrict}, a clear trend is observed. NSGA-II explores the widest range of solutions. It finds the lowest-card configurations, down to 67 cards, but with higher average delay. AGE-MOEA produces a smoother and more compact set of solutions. Its results show small variations in latency as the number of cards increases. CTAEA remains dominated in most of the feasible region.

Figure \ref{fig:front_max_restrict} shows the trade-off between the number of active cards and the maximum premium latency. All feasible solutions lie below the 7 ms threshold. Several configurations operate close to the limit, between 6.9 and 7.0 ms. This indicates that the optimization tends to use the full latency budget to reduce the number of active cards. Lower-card solutions are only possible when latency approaches the constraint.
NSGA-II again identifies the lowest-card feasible solutions, with 67 cards and maximum premium latency just below 7 ms. AGE-MOEA produces solutions concentrated between 74 and 87 cards, with maximum latency between 6.68 and 6.94 ms. CTAEA produces fewer competitive solutions, and some of its candidates exceed the 7 ms limit and are therefore not feasible (not shown in the table).

The combined non-dominated front highlights the best trade-offs under the constraint. As previously pointed out, the minimum feasible number of cards is 67, obtained with NSGA-II. This corresponds to a reduction of 25.6\% with respect to the 90-card Night Slice baseline. AGE-MOEA provides more stable solutions, with slightly higher resource usage but consistently controlled latency.

Compared with the unconstrained multi-service case, the feasible region is reduced. Solutions with fewer than 60 cards are no longer admissible, as they violate the 7 ms bound. In the previous case, such configurations provided the highest energy savings. These solutions are now excluded.

Despite this reduction, the framework still achieves significant savings while enforcing QoS guarantees. The results show that it is possible to maintain strict latency limits for premium traffic and still reduce the number of active cards by around 25\%.

The introduction of the premium latency constraint also affects the computational cost. For NSGA-II, the execution time increases from 770.73 s in the unconstrained case to 855.43 s under the 7 ms constraint (+11.0\%). A similar trend is observed for AGE-MOEA, where the execution time rises from 682.72 s to 756.66 s (+10.8\%). CTAEA shows a comparable behavior, increasing from 839 s to 936.05 s (+11.6\%). These results indicate that the constraint introduces a consistent overhead of around 11\% across all algorithms. This increase is due to the reduced feasible space, which makes it harder to identify valid solutions and requires more evaluations to converge. All experiments were conducted on an Apple M2 platform with 24 GB of RAM, using 6 CPU cores for parallel evaluation.

\section{Conclusion}
\label{conclusion}
This paper has shown that it is possible to generate several slices to cover traffic profiles at different times of the day or different days. A Day Slice is created following the traditional metrics, while a Night Slice is optimized for energy efficiency, keeping latency constraints within expected limits.

Results show that important savings can be achieved in terms of energy efficiency when selecting appropriate routing metrics. 
By using Pareto-based evolutionary algorithms, we demonstrate that by tuning link-weight we can effectively balance energy savings and network performance across diverse operating conditions.

In the single-service case, NSGA-II, CTAEA, and AGE-MOEA jointly minimized energy consumption and latency, revealing that up to 45\% of router line cards can be deactivated during low-traffic periods without compromising connectivity. AGE-MOEA achieved the most balanced trade-off between energy and delay, while CTAEA reached the lowest energy configurations at the expense of higher latency.

Building on these results, a multi-service extension  incorporated differentiated Quality of Service (QoS) constraints. By optimizing premium traffic maximum latency and cards, the framework produced a set of Pareto-optimal solutions that balance energy savings with strict latency guarantees for premium traffic. This demonstrates that energy-aware link-weight optimization can be made QoS-compliant and service-aware, supporting realistic multi-slice IP network scenarios.

Overall, the results confirm that Pareto-based multi-objective optimization provides a flexible foundation for adaptive network slicing.

Future work will explore modern services that rely on anycast or multi-destination routing (where traffic may be directed to any one of a group of candidate nodes). 

The current work demonstrates the concept in a time-exclusive dual-slice case (Day and Night) to clearly show the energy-saving potential. The framework itself does not preclude multiple slices running concurrently. In fact, the underlying IETF IP slicing model assumes that several logical slices can coexist on the same physical network.

\section*{Acknowledgments}
The authors would like to acknowledge the support of EU-funded ALLEGRO (grant No.101092766),  R\&D project PID2022-136684OB-C21 (Fun4Date) funded by the Spanish Ministry of Science and Innovation MCIN/AEI/10.13039/501100011033 and TUCAN6-CM (TEC-2024/COM-460), funded by CM (ORDEN 5696/2024)

\bibliographystyle{unsrt}  
\bibliography{references}

\end{document}